\documentclass[12pt, draftclsnofoot, onecolumn]{IEEEtran}
\usepackage{url,amsmath,graphicx}
\usepackage{fancyhdr}
\usepackage{cite}
\usepackage{graphicx}
\usepackage{psfrag}
\usepackage{subfigure}
\usepackage{url}
\usepackage{stfloats}
\usepackage{amsmath}
\usepackage{array}
\usepackage{fancyhdr}
\usepackage{epsfig}
\usepackage{amssymb}
\usepackage{color}
\newcommand\Exp{\mathbb{E}}
\usepackage{selinput}
\SelectInputMappings{%
	adieresis={ä},
	eacute={é},
	Lcaron={Ľ},
}

\usepackage[font=scriptsize]{caption}

\begin{document}
\title{ Mixed M\'alaga-$\mathcal{M }$ and Generalized-$\cal K$ Dual-Hop FSO/RF
Systems with Interference}

\author{ Im\`ene~Trigui, Member, \textit{IEEE}, Nesrine~Cherif, Sofi\`ene~Affes, Senior Member \textit{IEEE}, Xianbin Wang, Fellow, \textit{IEEE}, and Victor C. M. Leung, Fellow, \textit{IEEE}. }


\maketitle
\begin{abstract}
This paper  investigates the impact of  radio frequency (RF)
cochannel interference (CCI) on the performance of  dual-hop free-space optics (FSO)/RF
relay networks. The considered FSO/RF system operates over mixed  M\'alaga-$\mathcal{M}$/composite fading/shadowing generalized-$\cal K$ ($\cal GK$) channels with pointing errors.
The {\rm H}-transform theory, wherein integral transforms
involve Fox's {\rm H}-functions as kernels, is embodied into a unifying performance analysis framework  that encompasses closed-form expressions  for the outage probability, the average bit error rate (BER), and the ergodic capacity. By virtue of some {\rm H}-transform asymptotic expansions, the  high  signal-to-interference-plus-noise ratio (SINR) analysis culminates in easy-to-compute expressions for the outage probability and BER.
	\let\thefootnote\relax\footnotetext{Work supported by the Discovery Grants and the CREATE PERSWADE (www.create-perswade.ca) programs of NSERC, a Discovery Accelerator Supplement (DAS) Award from NSERC, and the NSERC SPG Project on Advanced Signal Processing and Networking Techniques for Cost-Effective Ultra-Dense 5G Networks.  This work
was presented at the IEEE PIMRC 2017.}
\end{abstract}
\section{Introduction}
Free-space optics (FSO) communication has recently drawn a significant attention as one promising solution  to cope with radio frequency (RF) wireless spectrum scarcity \cite{Cambridge}.
Though securing  high data rates, FSO communications performance  significantly degrades  due to atmospheric turbulence-induced
fading and strong path-loss \cite{Yang2}. Aiming  to address these shortcomings,   relay-assisted FSO systems have been actually identified as an influential solution to provide more efficient and wider networks.  As such, understanding the fundamental system performance  limits of mixed FSO/RF architectures has attracted a lot of research endeavor  in the past decade (cf.\cite{zedini},\!\!\cite{wcl} and references therein).

Up until recent past, the performance of relay-assisted FSO systems  was investigated assuming several  irradiance probability density function (PDF) models  with different
degrees of success out of which the most commonly utilized
models are the lognormal \cite{soleimani2016generalized} and the Gamma-Gamma\cite{emmna} PDFs. Recently, a new generalized statistical model, the M\'alaga-$\mathcal{M}$,  unifying most statistical models exploited so far and able to better reflect a wider range of turbulence conditions was  proposed in \cite{mal}, \cite{navas2}. Several performance studies of FSO link operating over M\'alaga-$\mathcal{M}$ turbulent channels with and without pointing errors have been conducted in \cite{wcl}, \cite{ansari}.

On the RF side, previous works typically assume either Nakagami-$m$ \cite{zedini}, \cite{emmna} or Rayleigh \cite{Yang}, \cite{ansari1} fading, thereby lacking the flexibility to account for disparate signal
propagation mechanisms as those characterized  in 5G communications which will accommodate a wide range of usage scenarios with diverse link requirements. In fact, in 5G communications
design, the combined effect of small-scale and shadowed fading needs to be properly addressed. Shadowing, which is due to obstacles in the local environment
or human body (user equipments) movements, can impact link performance by
causing fluctuations in the received signal.  For instance, the shadowing effect comes to prominence in millimeter wave
(mmWave) communications due to their higher carrier frequency. In this respect, the generalized-$\cal K$ ($\cal GK$) model was proposed by combining Nakagami-$m$ multipath
fading and Gamma-Gamma distributed shadowing \cite{imenet},\!\!\cite{miridakis2}.

While FSO transmissions are robust to RF interference, mixed FSO/RF systems are inherently vulnerable to the harmful effect of co-channel interference (CCI) through the RF link  (cf. \cite{imen2} and references therein). Previous contributions pertaining to FSO relay-assisted communications \cite{zedini}-\!\!\!\cite{ansari1}  relied on the absence of CCI. Recently, the recognition of the interference-limited nature  of emerging communication systems has motivated \cite{inter} to account for CCI in  the performance analysis of mixed decode and forward RF/FSO systems. Besides ignoring the shadowing effect on the RF link,  \cite{inter}  assumes a restrictive  Gamma-Gamma model on the FSO link.

In this paper, motivated by the aforementioned challenges, we assess
the impact of RF CCI on the performance of dual-hop amplify and forward (AF) mixed FSO/RF systems operating over M\'alaga-$\mathcal{M}$ and composite fading shadowing generalized-$\cal K$ ($\cal GK$)
channels, respectively.  Assuming fixed-gain and CSI-assisted relaying schemes and taking into account  the effect
of  pointing errors while considering both heterodyne and intensity modulation/direct (IM/DD) detection techniques,  we present a comprehensive
performance analysis by exploiting seminal results form the {\rm H}-transform theory. In addition, we present asymptotic
expressions for the outage probability and the average BER at high SINR and we
derive the diversity gain.

The remainder of this paper is organized as follows. We
describe the system model in Section II. In Section III, we present
 the unifying H-transform analysis of the end-to-end SINR statistics for both fixed-gain
and CSI-assisted relays.  Then, in section IV, we derive exact closed-form expressions for the outage probability,
the average BER, and the ergodic capacity followed by the
asymptotic expressions at high SINR. Section V presents some numerical and simulation results to
illustrate the mathematical formalism presented in the previous
sections. Finally, some concluding remarks are drawn out in
Section VI.

\section{Channel and System Models}
\label{sec:1} 
\begin{figure}[t!]
	\centering
	\includegraphics[scale=0.4]{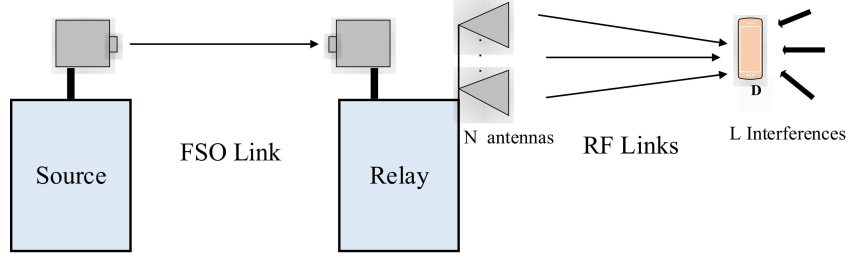}
	\caption{A dual-hop interference-limited mixed FSO/RF relay system.}
	\label{fig:fif}
\end{figure}

We consider a downlink of a relay-assisted network featuring a mixed FSO/RF communication. We assume that the optical source ($S$) communicates with the destination ($D$)  in a dual-hop fashion through an intermediate  relay ($R$). The latter is able to activate either heterodyne or IM/DD detection techniques at the reception of the optical beam. Using AF relaying, the relay  amplifies the received optical signal and
retransmits it to the destination with  MRT using $N$ antennas.  We assume that the destination is subject to inter-cell interference ($I$) brought by  $L$ co-channel RF sources in the network (cf. Fig.\ref{fig:fif}).

The optical ($S$-$R$) channel  follows a M\'alaga-$\mathcal{M}$  distribution for which the CDF of the  instantaneous SNR $\gamma_1$  in the presence of pointing errors is given  by
\begin{eqnarray}
	\label{eq:9}
	F_{\gamma_1}(x)&=&\frac{\xi^2A r}{\Gamma(\alpha)}\sum_{k=1}^{\beta}\frac{b_k}{\Gamma(k)}{\rm H}_{2,4}^{3,1}\Biggl[\frac{B^rx}{\mu_r}\Bigg\vert{ (1,r),(\xi^2+1,r) \atop(\xi^2,r),(\alpha,r),(k,r),(0,r)}\Biggr],
\end{eqnarray}
where  $\xi$ is the ratio between the equivalent beam radius and the pointing error displacement standard deviation (i.e., jitter) at the relay (for negligible pointing errors $\xi \rightarrow+\infty$) \cite{Yang2},
$ A={ \alpha^{\frac{\alpha}{2}}\left[  {g \beta }/({g \beta +\Omega})\right] ^{\beta +\frac{\alpha}{2}}}{g^{-1-\frac{\alpha}{2}}}$ and $ {b_k}\!\!=\!\binom{\beta\!-\!1}{k\!-\!1}\!{\left(g \beta\!+\Omega \right)^{1-\frac{k}{2}}}\left[ {(g \beta +\!\Omega)}/{\alpha\beta}\right] ^{\frac{\alpha+k}{2}}\left( {\Omega}/{g}\!\right)^{k-1}\left({\alpha}/{\beta}\!\right)^{\frac{k}{2}}$, where $\alpha$, $\beta$, $g$  and $\Omega$ are the fading parameters related to the atmospheric turbulence conditions \cite{ansari}. It may be useful to mention that $g=2b_{0}(1-\rho)$ where $2 b_{0}$ is the average power of the LOS term and $\rho$ represents the amount of scattering power coupled to the LOS component ($0  \leqslant \rho_i  \leqslant 1 $). Moreover in (\ref{eq:9}), ${\rm H}_{p, q}^{m, n} [\cdot]$ and $\Gamma(\cdot)$ stand for the Fox-{\rm H} function \cite[Eq.(1.2)]{mathai} and the incomplete gamma function \cite[Eq.(8.310.1)]{grad}, respectively, and $B={\alpha\beta h (g+\Omega)}/[{(g\beta+\Omega)}]$ with  $h=\xi^2/(\xi^2+1)$. Furthermore, $r$ is the parameter that describes the detection technique at the relay (i.e., $r = 1$
is associated with heterodyne detection and $r = 2$ is associated
with IM/DD) and, $\mu_r$ refers to the electrical SNR of the FSO
hop \cite{ansari}. In particular, for $r = 1$,
\begin{equation}
\mu_{1}=\mu_{\text{heterodyne}}=\Exp[\gamma_1]=\bar{\gamma}_1,
\end{equation}
and for $r=2$,  it becomes \cite[Eq.(8)]{ansari}
\begin{equation}
\mu_{2}=\mu_{\text{IM/DD}}= \frac{\mu_{1}\alpha\xi^2(\xi^2+1)^{-2}(\xi^2+2)(g+\Omega)}{(\alpha+1)[2g(g+2\Omega)+\Omega^2(1+\frac{1}{\beta})]}.
\end{equation}

The RF ($R$-$D$) and ($I$-$D$) links are assumed to follow generalized-$\cal K$  fading distributions.  Hence the probability density function (PDF) of the instantaneous SNR (respectively  INR), $\gamma_{XD}$,  $X\in (R, I)$, is  given by \cite[Eq.(5)]{imenet}
\begin{eqnarray}
	\label{eqC2:4}
	f_{\gamma_{XD}}(x)&=&\frac{2\left( \frac{ m_X\kappa_X}{\bar{\gamma}_{XD}}\right)^{\frac{\kappa_X+\delta_Xm_X}{2}}x^{\frac{\kappa_X+\delta_Xm_X}{2}-1} }{\Gamma(\delta_Xm_X)\Gamma(\kappa_X)}{ K}_{\kappa_X-\delta_Xm_X}\left( 2\sqrt{\frac{\kappa_Xm_X x}{\bar{\gamma}_{XD}}}\right),
\end{eqnarray}
where $X\in\{R,I\}$ and $K_\nu(\cdot)$ stands for the modified Bessel function of the second kind \cite[Eq.(8.407.1)]{grad}. Moreover, $m_X\geqslant	0.5$ and $\kappa_X\geqslant 0$ denote the
multipath fading and shadowing severity of the $X$-$D$th channel coefficient, respectively. Moreover, $\delta_X=\{N,L\}$ for $X\in \{R,I\}$ follows form the conservation property under the  summation of  $N$ and $L$ i.i.d . (independent identically distributed) $\cal GK$ random variables. The interfering signals are assumed to propagate through i.i.d $\cal GK$ channels
with parameters $m_I$ and $\kappa_I$. Using \cite[Eq.(9.34.3)]{grad}, the PDF of the $\cal GK$ distribution can be represented in terms of the Meijer's-G function as
\begin{eqnarray}
	\label{eqC2:4bis}
	f_{\gamma_{XD}}(x)&=&\frac{\frac{m_X \kappa_X}{\bar{\gamma}_{XD}}}{\Gamma(\delta_Xm_X)\Gamma(\kappa_X)}{\rm G}_{0,2}^{2,0}\Biggl[\!\frac{\kappa_X\!m_X}{\bar{\gamma}_{XD}} x\Bigg\vert \  {-\atop \delta_Xm_X\!-\!1,\kappa_X-1}\Biggr].
\end{eqnarray}
The CDF of the signal-to-interference ratio (SIR)  $\gamma_2=\gamma_{RD}/\gamma_{ID}$ under $\cal GK$ fading can be derived from a recent result in \cite[Lemma 1]{miridakis2} as
\begin{eqnarray}
	\label{eqC2:5}
	F_{\gamma_2}(x)&=&1-\frac{1}{\Gamma(Nm)\Gamma(\kappa)\Gamma(Lm_I)\Gamma(\kappa_I)}{\rm G}_{3,3}^{3,2}\Biggl[\frac{\kappa m x}{\kappa_Im_I\bar{\gamma}_2}\Bigg\vert  {1-\kappa_I,1-Lm_I,1 \atop 0,\kappa,Nm}\Biggr],
\end{eqnarray}
where  $\bar{\gamma}_2=\bar{\gamma}_{RD}/\bar{\gamma}_{ID}$ is the average SIR of the RF link where, for consistency, we have dropped the subscript $R$ from the parameters $m_R$ and $\kappa_R$.

In the fixed-gain relaying scheme, the end-to-end SINR at the destination can be expressed as \cite[Eq.(2)]{imenefixed}
\begin{equation}
	\label{eqC2:8}
	\gamma=\frac{\gamma_1\gamma_2}{\gamma_2 +{\cal C}},
\end{equation}
where $\cal C$ stands for the fixed gain at the relay. Whereas, the end-to-end SINR when CSI-assisted relaying scheme is considered is
 expressed  as \cite[Eq.(7)]{Yang}
 \begin{equation}
 \label{eq:10}
 \gamma=\frac{\gamma_1 \gamma_2}{\gamma_1+\gamma_2+1}.
 \end{equation}
\section{End-to-End Statistics}
\label{sec:2}

\subsection{Fixed-Gain Relaying }
The CDF of the end-to-end SINR of interference-limited dual-hop FSO/RF systems using a fixed-gain relay in M\'alaga-$\mathcal{M}$/$\cal GK$ fading under both heterodyne detection and IM/DD is given  by
\begin{eqnarray}
	\label{eqC2:CDFfix}
	\!\!F_\gamma(x)&\!\!=\!\!&\frac{\xi^2A \kappa m {\cal C} }{\Gamma(\alpha)\Gamma(Nm)\Gamma(\kappa)\Gamma(Lm_I)\Gamma(\kappa_I)\kappa_Im_I\bar{\gamma}_2}\sum_{k=1}^{\beta}\frac{b_k}{\Gamma(k)}{\rm H}_{1,0:3,2:4,5}^{0,1:0,3:4,3}\left[{ \frac{\mu_r}{B^rx}\atop \frac{\kappa m{\cal C}}{\kappa_Im_I\bar{\gamma}_2}} \left|\begin{array}{cccc}(0,1, 1) \\-\\(\delta,\Delta)\\(\lambda,\Lambda)\\(\chi, X)\\(\upsilon,\Upsilon) \end{array}\right.\right],
\end{eqnarray}
where  ${\rm H}^{m_1,n_1:m_2,n_2:m_3,n_3}_{p_1,q_1:p_2,q_2:p_3,q_3}[\cdot]$ denotes the Fox-H function (FHF) of two variables\cite[Eq.(1.1)]{mittal} also known as the bivariate FHF whose Mathematica implementation may be found in \cite[Table I]{Lei}, whereby $(\delta,\Delta)=(1-\xi^2\!,r),(1-\alpha,r),(1-k,r)$; $(\lambda,\Lambda)=(0,1),(-\xi^2,r)$; $(\chi, X)=(-1,1),(-\kappa_I,1),(-Lm_I,1),(0,1)$; and $(\upsilon,\Upsilon)=(-1,1),(-1,1),(\kappa-1,1),(Nm-1,1),(0,1)$
\begin{IEEEproof}
See Appendix \ref{app:A}.
\end{IEEEproof}
The PDF of the end-to-end SINR $\gamma$ in mixed M\'alaga-$\mathcal{M}$/$\cal GK$ is obtained  as
\begin{eqnarray}
	\label{eqC2:15}
	f_\gamma(x)&=&-\frac{\xi^2A \kappa m {\cal C}  }{x\Gamma(\alpha)\Gamma(Nm)\Gamma(\kappa)\Gamma(Lm_I)\Gamma(\kappa_I)\kappa_Im_I\bar{\gamma}_2}\sum_{k=1}^{\beta}\frac{b_k}{\Gamma(k)}\nonumber \\ && {\rm H}_{1,0:3,2:4,5}^{0,1:0,3:4,3}\left[{ \frac{\mu_r}{B^rx}\atop \frac{\kappa m{\cal C}}{\kappa_Im_I\bar{\gamma}_2}} \left|\begin{array}{cccc}(0,1, 1) \\-\\(\delta,\Delta)\\(\lambda',\Lambda')\\(\chi, X)\\(\upsilon,\Upsilon) \end{array}\right.\right],
\end{eqnarray}	
where $(\lambda',\Lambda')=(1,1),(-\xi^2,r)$.
\begin{IEEEproof}
The result follows from  differentiating  the Mellin-Barnes integral in (\ref{eqC2:CDFfix}) over $x$ using $\frac{dx^{-s}}{dx}=-sx^{-s-1}$ with $\Gamma(s+1)=s\Gamma(s)$ and applying \cite[Eq.(2.57)]{mathai}.
\end{IEEEproof}
\subsection{CSI-Assisted Relaying }
 Due to the  intractability  of the SINR in (\ref{eq:10}), we resort to an upper bound given by \cite[Eq.(20)]{Yang} as $\gamma=\min(\gamma_1,\gamma_2)>\gamma_1\gamma_2/(\gamma_1+\gamma_2+1)$, whose CDF can be expressed as $F_\gamma(x)=1-F^{(c)}_{\gamma_1}(x)F^{(c)}_{\gamma_2}(x)$, where $F^{(c)}_{\gamma_1}$ and $F^{(c)}_{\gamma_2}$ stand for the complementary CDF of $\gamma_1$ and $\gamma_2$, respectively. Hence, using \cite[Eq.(8)]{wcl} and (\ref{eqC2:5}), the CDF of dual-hop FSO/RF systems employing a CSI-assisted relaying scheme can be obtained as
 \begin{eqnarray}
 \label{eq:CSICDF}
 F_{\gamma}(x)&=&1-\frac{\xi^2A}{\Gamma(\alpha)\Gamma(Nm)\Gamma(\kappa)\Gamma(Lm_I)\Gamma(\kappa_I)}\sum_{k=1}^{\beta}\frac{b_k}{\Gamma(k)}{\rm  G}_{2,4}^{4,0} \Biggl[B \left(\frac{x}{\mu_r}\right)^{\frac{1}{r}} \Bigg\vert \  {\xi^2+1,1 \atop 0,\xi^2,\alpha,k}\Biggr]\nonumber \\ && {\rm G}_{3,3}^{3,2}\!\Biggl[\frac{\kappa m x}{\kappa_Im_I\bar{\gamma}_2}\Bigg\vert  {1-\kappa_I,1-Lm_I,1 \atop 0,\kappa,Nm}\Biggr].
 \end{eqnarray}
\section{Performance Analysis of Fixed-Gain Relaying}
\label{sec:3}
\subsection{Outage Probability}
 The quality of service (QoS) of the considered mixed FSO/RF system  is ensured by keeping the instantaneous end-to-end SNR, $\gamma$, above a threshold $\gamma_{th}$.
	The outage probability of the considered mixed FSO/RF system follows from (\ref{eqC2:CDFfix}) as
	\begin{equation}
		\label{eqC2:outfix1}
		P_{\text{out}}=F_\gamma(\gamma_{th}).
	\end{equation}
\label{cor:1}
	At high normalized average SNR  in the FSO link ($\frac{\mu_r}{\gamma_{th}}\rightarrow\infty$), the outage probability of the system under consideration is obtained as
	\begin{eqnarray}
		\label{eqC2:outfix1high}
		P_{\text{out}}&\underset{\frac{\mu_r}{\gamma_{th}}\gg1}{\approx}&\frac{\xi^2A \frac{\kappa m}{\kappa_Im_I} {\cal C}}{\Gamma(\alpha)\Gamma(Nm)\Gamma(\kappa)\Gamma(Lm_I)\Gamma(\kappa_I) \bar{\gamma}_2}\sum_{k=1}^{\beta} \frac{b_k}{\Gamma(k)} \Biggl(\frac{\Gamma(\alpha-\xi^2)\Gamma(k-\xi^2)}{r \Gamma(1-\frac{\xi^2}{r})}\quad\Xi\left(\gamma_{th},\frac{\xi^2}{r}\right) \nonumber\\
		&&+\frac{\Gamma(\xi^2-\alpha)\Gamma(k-\alpha)}{r \Gamma(1-\frac{\alpha}{r})\Gamma(1+\xi^2-\alpha)}\quad\Xi\left(\gamma_{th},\frac{\alpha}{r}\right) +\frac{\Gamma(\xi^2-k)\Gamma(\alpha-k)}{r \Gamma(1-\frac{k}{r})\Gamma(1+\xi^2-k)}\quad\Xi\left(\gamma_{th},\frac{k}{r}\right) \nonumber\\
		&&+\frac{B^r\gamma_{th}}{\mu_r}{\rm H}_{6,8}^{7,3}\Biggl[{\frac{\kappa m{\cal C}B^r\gamma_{th}}{\kappa_Im_I\bar{\gamma}_2\mu_r}}\left|\begin{array}{cccc}(\sigma,\Sigma)\\(\phi,\Phi)\end{array}\right.\!\!\! \Biggr]\!\Biggr),
	\end{eqnarray}
	where
 \begin{equation}
 \Xi(x,y)=\left(\frac{B^rx}{\mu_r}\right)^{y}\!\!{\rm G}_{5,5}^{4,4}\!\!\left[\!\!{\frac{\kappa m{\cal C}}{\kappa_I m_I\bar{\gamma}_2}}\!\!\left|  \begin{array}{cccc}-\kappa_I\!,-\!Lm_I\!,\!-1,y,0\!\\ \kappa\!-\!1,Nm-1,\!-1,\!-1,0\end{array}\!\!\right.\right], \end{equation}  	$(\sigma,\Sigma)=(-\kappa_I,1),(-Lm_I,1),(-1,1),(0,1),(1+\xi^2-r,r),(0,1)$, and $(\phi,\Phi)=(\xi^2-r,r),(\alpha-r,r),(k-r,r),(\kappa-1,1),(Nm-1,1),(-1,1),(-1,1),(0,1)$.
\begin{IEEEproof}
	Resorting to the Mellin-Barnes representation of the bivariate FHF \cite[Eq.(2.57)]{mathai} in (\ref{eqC2:CDFfix}) and applying \cite[Theorem 1.7]{kilbas} yield (\ref{eqC2:outfix1high}) after some additional algebraic manipulations.
\end{IEEEproof}
Furthermore, when $\bar{\gamma}_2\rightarrow\infty$, then  by applying \cite[Theorem 1.11]{kilbas} to (\ref{eqC2:outfix1high}) while only keeping the dominant term,   the diversity gain for FSO/RF systems  with pointing errors  over  M\'alaga-$\mathcal{M}$/$\cal GK$ fading conditions can be shown to be equal to
 \begin{equation}
  {\rm G}_d=\min\left(N m, \kappa, \frac{\xi^2}{r},\frac{\alpha}{r},\frac{k}{r}\right).
  \label{div}
  \end{equation}
  In particular, under Nakagami-$m$ fading, i.e., when $\kappa\rightarrow\infty $,  we obtain    $ {\rm G}_d=\min\left(N m, \frac{\xi^2}{r},\frac{\alpha}{r},\frac{k}{r}\right)$ \cite[Eq. 29]{zedini}.
\subsection{Average Bit-Error Rate}
The average error probability for the considered
 dual-hop mixed RF/FSO AF relay system
with interference at the destination and pointing errors  at the FSO link under both heterodyne and IM/DD detection techniques is
analytically derived as
\begin{eqnarray}
		\label{eqC2:fix5}
		\overline{P}_e &=& \frac{\xi^2A \varphi\kappa m {\cal C}}{2\Gamma(\alpha) \Gamma(p)\Gamma(Nm)\Gamma(\kappa)\Gamma(Lm_I)\Gamma(\kappa_I)\kappa_Im_I \bar{\gamma}_2}\sum_{j=1}^{n}\sum_{k=1}^{\beta}\frac{b_k}{\Gamma(k)}\nonumber \\ &&{\rm H}_{1,0:3,3:4,5}^{0,1:1,3:4,3}\left[{ \frac{\mu_rq_j}{B^r}\atop \frac{\kappa m{\cal C}}{\kappa_Im_I\bar{\gamma}_2}} \left|\begin{array}{cccc}(0,1, 1) \\-\\(\delta,\Delta)\\(p,1),(\lambda,\Lambda)\\(\chi, X)\\(\upsilon,\Upsilon) \end{array}\right.\right].
	\end{eqnarray}	
\begin{IEEEproof}
The average BER
can be written in terms of the CDF of the end-to-end SINR  as
\begin{equation}\label{eqC2:ber1}
	\overline{P}_e=\frac{\varphi}{2\Gamma(p)}\sum_{j=1}^{n}q_j^p\int_{0}^{\infty}e^{-q_jx}x^{p-1}F_{\gamma}(x)\mathrm{d}x,
\end{equation}
where $\Gamma(\cdot,\cdot)$ stands for the incomplete Gamma function \cite[Eq.(8.350.2)]{grad} and the parameters $\varphi$, $n$, $p$ and $q_j$ account for different modulations schemes \cite{imenet}.
Now, substituting the Mellin-Barnes integral form  of (\ref{eqC2:CDFfix}) using \cite[Eq.(2.56)]{mathai} into (\ref{eqC2:ber1}), and resorting to \cite[Eq.(7.811.4)]{grad} yield (\ref{eqC2:fix5}) after some manipulations.
\end{IEEEproof}	
	At high FSO SNR (i.e. $\mu_r \rightarrow\infty$), the asymptotic average BER  is derived  as
	\begin{eqnarray}
		\label{eqC2:berfix1high}
		\overline{P}_e\!&\underset{\mu_r\gg1}{\approx}&\!\!\!\!\frac{\xi^2A \varphi\kappa m {\cal C}}{2\Gamma(\alpha) \Gamma(p)\Gamma(Nm)\Gamma(\kappa)\Gamma(Lm_I)\Gamma(\kappa_I)\kappa_Im_I \bar{\gamma}_2}\sum_{j=1}^{n}\sum_{k=1}^{\beta}\frac{b_k}{\Gamma(k)}\Biggl[\frac{\Gamma(\alpha-\xi^2)\Gamma(k-\xi^2)}{r \Gamma(1-\frac{\xi^2}{r})}\Xi\left(\frac{1}{q_j}\frac{\xi^2}{r}\right) \nonumber\\
		&&\!\!\!\!+\frac{\Gamma(\xi^2-\alpha)\Gamma(k-\alpha)}{r \Gamma(1-\frac{\alpha}{r})\Gamma(1+\xi^2-\alpha)}\quad\Xi\left(\frac{1}{q_j},\frac{\alpha}{r}\right)+\frac{\Gamma(\xi^2-k)\Gamma(\alpha-k)}{r \Gamma(1-\frac{k}{r})\Gamma(1+\xi^2-k)}\quad\Xi\left(\frac{1}{q_j},\frac{k}{r}\right) \nonumber\\
		&&\!\!\!\!+\frac{B^r}{\mu_rq_j}{\rm H}_{7,8}^{7,4}\Biggl[{\frac{\kappa m{\cal C}B^r}{\kappa_Im_I\bar{\gamma}_2\mu_rq_j}}\left|\begin{array}{cccc}(\sigma',\Sigma')\\(\phi,\Phi)\end{array}\right.\!\!\! \Biggr]\!\Biggr],
	\end{eqnarray}
	where $(\sigma',\Sigma')=(-\!\kappa_I,1),(-\!Lm_I,1),(-1,1),(-p,1),(0,1),(1+\xi^2-r,r),(0,1)$.
	\begin{IEEEproof}
		The asymptotic BER follows  along the same lines  as (\ref{eqC2:outfix1high}).
	\end{IEEEproof}
\subsection{Ergodic Capacity}
	The ergodic capacity of a mixed M\'alaga-$\mathcal{M}$/interference-limited $\cal GK$ transmission system under both detection techniques with pointing errors at the FSO link  is obtained as
	\begin{eqnarray}
		\label{eqC2:fix21}
\overline{C}&=&\frac{\xi^2A \kappa m {\cal C} }{2 \ln(2)\Gamma(\alpha)\Gamma(Nm)\Gamma(\kappa)\Gamma(Lm_I)\Gamma(\kappa_I)\kappa_Im_I\bar{\gamma}_2}\sum_{k=1}^{\beta} \frac{b_k}{\Gamma(k)}\nonumber \\ && {\rm H}_{1,0:4,3:4,5}^{0,1:1,4:4,3}\left[{ \frac{\mu_r}{B^rx}\atop \frac{\kappa m{\cal C}}{\kappa_Im_I\bar{\gamma}_2}}\left|\begin{array}{cccc}(0,1, 1) \\-\\(\delta,\Delta),(1,1)\\(0,1)(\lambda',\Lambda')\\(\chi, X)\\(\upsilon,\Upsilon) \end{array}\right.\right].
	\end{eqnarray}	  	 	 	 	
\begin{IEEEproof}
	The ergodic capacity $\overline{C}=\frac{1}{2}\mathbb{E}\left[\ln_2(1+\gamma)\right]$ follows from averaging $\ln(1+\gamma)={\rm G}_{2,2}^{1,2} [ \gamma \vert\!\!\ {{1,1} \atop {1,0}}]$ over the end-to-end SINR PDF obtained in (\ref{eqC2:15}) while resorting to \cite[Eq.(1.1)]{mittal} and \cite[Eq.(7.811.4)]{grad}  with some  manipulations.
\end{IEEEproof}
The M\'alaga-$\mathcal{M}$ reduces to Gamma-Gamma fading when  ($g=0$, $\Omega=1$), whence  all terms in (\ref{eq:9}) vanish except for
the term when  $k=\beta$.
Hence, when $g=0$, $\Omega=1$, $\kappa, \kappa_I \rightarrow \infty$, (\ref{eqC2:fix21}) reduces, when $r=1$,  to the ergodic capacity of mixed Gamma-Gamma FSO/interference-limited Nakagami-$m$ RF transmission with heterodyne detection as given by
	\begin{eqnarray}
		\label{eqC2:fixGGN}
		\overline{C}&=&\frac{\xi^2}{2 \ln(2)\Gamma(Nm)\Gamma(Lm_I)\Gamma(\alpha)\Gamma(\beta)}\nonumber\\ && {\rm G}_{1,0:4,3:4,3}^{1,0:1,4:3,2}\Biggl[\frac{\mu_1}{\alpha\beta h}; \frac{ m{\cal C}}{m_I\bar{\gamma}_2}\Bigg\vert\ {1\atop-}\ \Bigg\vert\ {1-\xi^2,1-\alpha,1-\beta,1\atop 1,0,-\xi^2}\!\Bigg\vert \ {1-Lm_I,1,0\atop Nm,0,1}\Biggr],\nonumber\\
	\end{eqnarray}
where ${\rm G}_{a,[c,e],b,[d,f]}^{p,q,k,r,l}[\cdot,\cdot]$  is   the generalized Meijer's G-function and is used to represent the product of three Meijer's-G
functions in a closed-form  \cite{verma}.
\section{Performance Analysis of CSI-assisted relaying}
\subsection{Outage Probability}
Based on (\ref{eq:CSICDF}), the outage probability of CSI-assisted mixed M\'alaga-$\mathcal{M}$ turbulent/$\cal GK$  systems with interference under both detection techniques with pointing errors can be lower bounded by
\begin{eqnarray}
\label{eq:poutsci}
P_{\text{out}}^{\text{lb}}&=&1-\frac{\xi^2Ar}{\Gamma(\alpha)\Gamma(Nm)\Gamma(\kappa)\Gamma(Lm_I)\Gamma(\kappa_I)}\sum_{k=1}^{\beta}\frac{b_k}{\Gamma(k)}{\rm H}_{0,0:2,4:3,3}^{0,0:4,0:3,2} \left[{ \frac{B^r\gamma_{th}}{\mu_r}\!\atop \frac{\kappa m\gamma_{th}}{\kappa_Im_I\bar{\gamma}_2}} \!\left|\begin{array}{cccc}(0,1, 1\!) \\-\\(\delta_1,\Delta_1)\\(\lambda_1,\Lambda_1)\\(\chi_1, X_1)\\(\upsilon_1,\Upsilon_1) \end{array}\right.\right],
\end{eqnarray}
where $(\delta_1,\Delta_1)=(\xi^2+1,r),(1,r)$, $(\lambda_1,\Lambda_1)=(0,r),(\xi^2,r),(\alpha,r),(k,r)$, $(\chi_1, X_1)=(1-\kappa_I,1),(1-Lm_I,1),(1,1)$, and $(\upsilon_1,\Upsilon_1)=(0,1),(\kappa,1),(Nm,1)$.
\subsection{Average Bit-Error Rate}
	The average BER of a mixed FSO/interference-limited RF CSI-assisted relaying system in  M\'alaga-$\mathcal{M}$ turbulent with pointing errors/$\cal GK$  fading channels  under both detection techniques is obtained as
	\begin{eqnarray}
	\label{eq:17}
	\overline{P_e}&=&\frac{\varphi n}{2}-\frac{ \xi^2A r\varphi
	}{2\Gamma(p)\Gamma(\alpha)\Gamma(Nm)\Gamma(\kappa)\Gamma(Lm_I)\Gamma(\kappa_I)}\nonumber \\ &&\sum_{j=1}^{n} \sum_{k=1}^{\beta}\frac{b_k}{\Gamma(k)} {\rm H}_{1,0:2,4:3,3}^{0,1:4,0:3,2}\left[\!\!{\frac{B^r}{\mu_rq_j}\atop \frac{\kappa m}{\kappa_Im_I\bar{\gamma}_2q_j}}\left| \begin{array}{cccc}(1-p,1, 1) \\-\\(\delta_1,\Delta_1)\\(\lambda_1,\Lambda_1)\\(\chi_1, X_1)\\(\upsilon_1,\Upsilon_1) \end{array}\!\!\right.\right].
	\end{eqnarray}  	
\begin{IEEEproof}
	Substituting (\ref{eq:CSICDF}) into (\ref{eqC2:ber1}) and resorting to  \cite[Eq.(1.59)]{mathai} and \cite[Eq.(2.2)]{mittal} yield the result  after some manipulations.
\end{IEEEproof}
\subsection{Ergodic Capacity}
	The ergodic capacity of a mixed FSO/interference-limited RF CSI-assisted relaying system  in  M\'alaga-$\mathcal{M}$/$\cal GK$  fading channels under both detection techniques is expressed by
	\begin{eqnarray}
	\label{eq:18}
\!\!\!\!\!\overline{C}&\!\!=\!\!&\frac{\xi^2A  r  \mu_r}{2 \ln(2)\Gamma(\alpha)\Gamma(Nm)\Gamma(\kappa)\Gamma(Lm_I)\Gamma(\kappa_I)B^{r}} \sum_{k=1}^{\beta} \frac{b_k}{\Gamma(k)} {\rm H}_{1,0:4,3:3,4}^{0,1:1,4:3,3} \left[{\frac{\mu_r}{B^r}\!\atop \frac{\kappa_Im_I\bar{\gamma}_2}{\kappa m}} \left| \begin{array}{cccc}(0,1, 1) \\-\\(\delta_2,\Delta_2)\\(\lambda_2,\Lambda_2)\\(\chi_2, X_2)\\(\upsilon_2,\Upsilon_2) \end{array}\right.\right],
	\end{eqnarray}
	where $(\delta_2,\Delta_2)=(1\!-r\!,r\!),(1\!-\xi^2\!-r,\!r)\!, \!(1\!-\!\alpha\!-r,\!r)\!,\!(1\!-\!k\!-r,\!r)\!$, $(\lambda_2,\Lambda_2)=(1,1),(\!1-\kappa,1),(\!1-\!Nm,1)$, $(\chi_2, X_2)=(1,1),(\!1-\kappa,1),(\!1-\!Nm,1)$, and $(\upsilon_2,\Upsilon_2)=(1,1),(\kappa_I,1),(Lm_I,1),(0,1)$.
\begin{IEEEproof}
	See Appendix \ref{app:B}.
\end{IEEEproof}
	It should be mentioned that when $r=1$ and  $\kappa, \kappa_I \rightarrow \infty$,  (\ref{eq:18})  reduces to  the ergodic capacity of mixed FSO/interference-limited RF systems in M\'alaga/Nakagami-$m$ fading channels   as given by
	\begin{eqnarray}
	\label{eq:15bis}
	\overline{C}&=&\frac{\xi^2A   \mu_1}{2 \ln(2)B\Gamma(\alpha)\Gamma(Nm)\Gamma(Lm_I)\alpha\beta h}\sum_{k=1}^{\beta} \frac{b_k}{\Gamma(k)}\nonumber \\ && {\rm G}_{1,0:4,3:2,3}^{1,0:1,4:2,2}\Biggl[\frac{\mu_1}{\alpha\beta h}; \frac{m_I\bar{\gamma}_2}{m}\Bigg\vert \ {1\atop-}\ \Bigg\vert\ {0,-\xi^2,-\alpha,-k\atop 0,-\xi^2-1,-1}\!\Bigg\vert \  {1,1-Nm\atop 1,Lm_I,0}\Biggr].
	\end{eqnarray}

\section{Numerical results}
\label{sec:4}
In this section, numerical examples are shown to substantiate
the accuracy of the new unified mathematical framework and to confirm
its potential for analyzing mixed FSO/RF communications. Remarkably, all numerical results obtained by the direct evaluation of the analytical expressions developed
in this paper,  are in very good match with their Monte-Carlo stimulated counterparts showing  the accuracy and effectiveness of our new performance analysis framework.
 Unless stated otherwise, all  simulations were carried out with the following parameters:  ${\cal C}=1.7$,  $m_I=1.5$,  $\kappa_I=3.5$, and $\bar{\gamma}_2=20$ dB.

\begin{figure}[!t]
	\centering\begin{minipage}{.6\textwidth}
	\includegraphics[width=\textwidth]{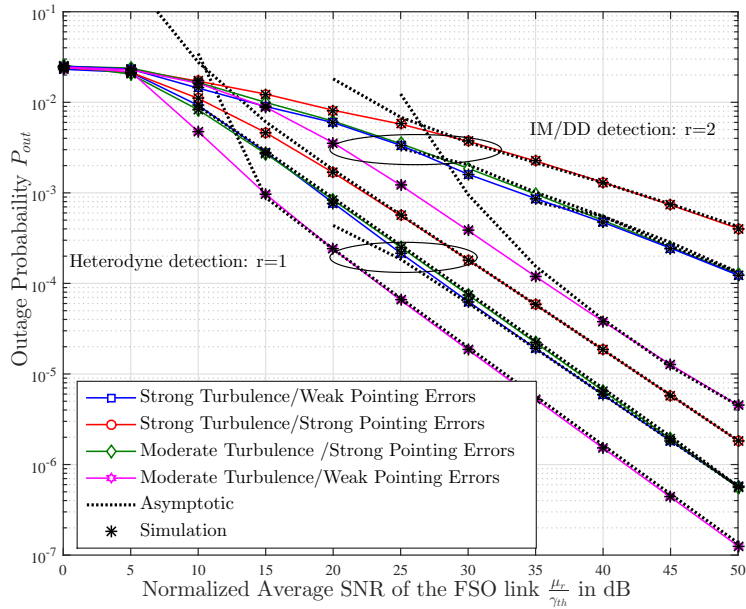}
	\caption{Outage probability of a fixed-gain mixed RF/FSO system with interference under different turbulence and pointing errors severities  with $N=L=2$, $m=2.5$, and $\kappa=1.09$.}
	\label{fig:outfixed}\end{minipage}
\end{figure}

Fig.~\ref{fig:outfixed} illustrates the outage probability of mixed FSO/RF  fixed-gain AF systems versus the  FSO link normalized average SNR in strong  (i.e., $\alpha=2.4$, $\beta=2$) and weak (i.e., $\alpha=5.4$, $\beta=4$) turbulence conditions, respectively. The figure also investigates the effect of strong (i.e., $\xi=1.1$) and  weak (i.e., $\xi=6.8$) pointing errors on the system performance.
As expected, the outage probability deteriorates by decreasing the pointing error displacement standard deviation, i.e., for smaller $\xi$, or decreasing the turbulence fading parameter, i.e., smaller $\alpha$ and $\beta$. At high SNR, the asymptotic expansion in (\ref{eqC2:outfix1high}) matches very well its exact counterpart, which confirms the validity of our mathematical analysis for different parameter settings. On the other hand, we observe that  heterodyne detection outperforms IM/DD in turbulent environments as previously observed in \cite{ansari}.

\begin{figure}[!t]
	\centering\begin{minipage}{.6\textwidth}
	\includegraphics[width=\textwidth]{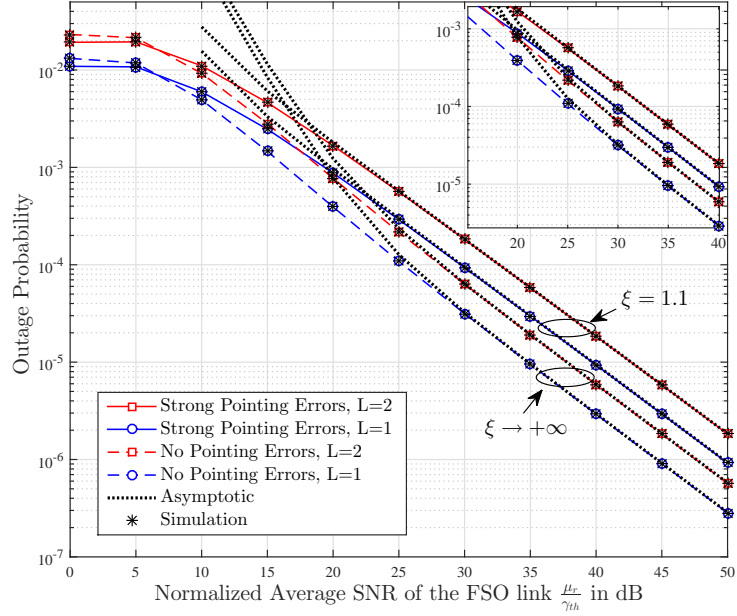}
	\caption{Outage probability of an interference-limited  fixed-gain mixed RF/FSO system in strong turbulence conditions for different values of  $L$ and $\xi$  with $N=L=2$, $m=2.5$, and $\kappa=1.09$.}
	\label{fig:outfixedinterference}\end{minipage}
\end{figure}

 Fig.~\ref{fig:outfixedinterference} depicts the outage probability of fixed-gain   mixed FSO/interference-limited RF systems with $L=\{1,2\}$ versus  the  FSO link normalized average SNR. As expected, increasing $L$ deteriorates the system performance, by increasing the outage probability while the diversity gain remains unchanged. Once again we highlight the fact  that the exact and asymptotic expansion in  (\ref{eqC2:outfix1high}) agree very well at high SNRs.
\begin{figure}[!t]
	\centering\begin{minipage}{.6\textwidth}
	\includegraphics[width=\textwidth]{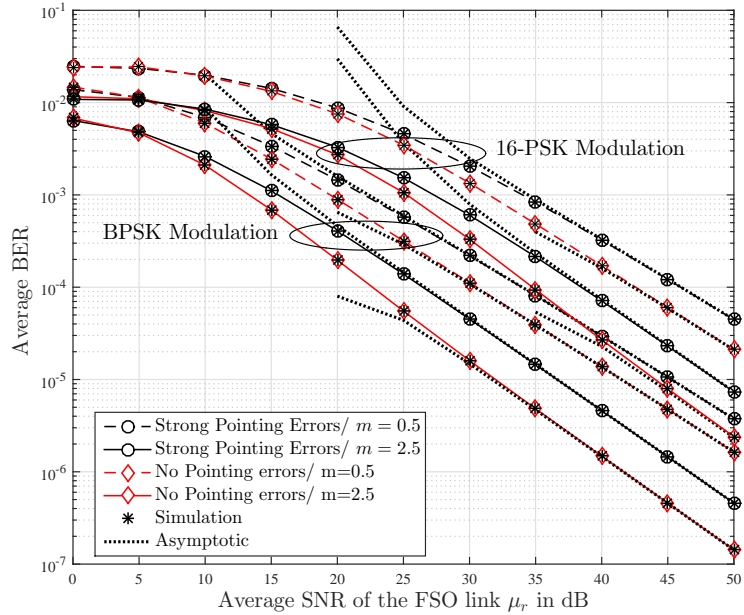}
	\caption{Average BER  of an interference-limited  fixed-gain mixed RF/FSO system in strong turbulence conditions for different values of $m$  with $N=L=2$,  and $\kappa=1.09$.}
	\label{fig:berfixed}\end{minipage}
\end{figure}

\begin{figure}[!t]
	\centering
\begin{minipage}{.6\textwidth}
	\includegraphics[width=\textwidth]{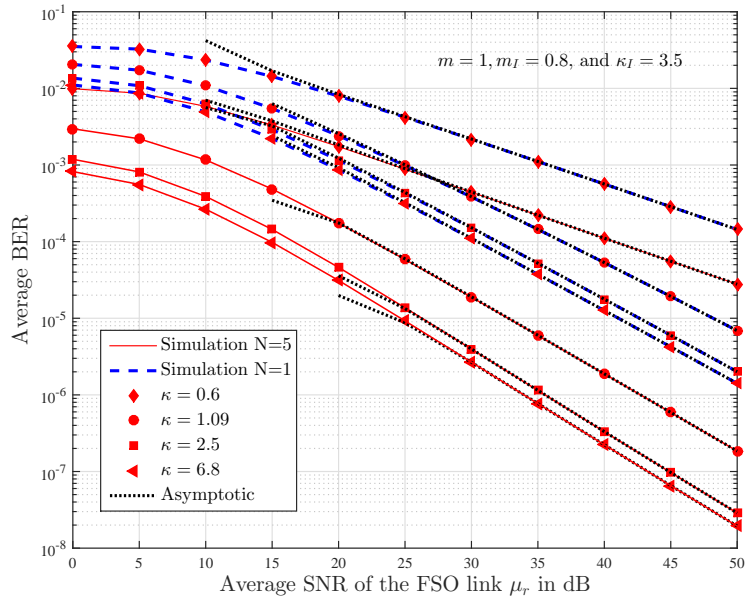}
	\caption{Average BER  of an interference-limited  fixed-gain mixed FSO/RF system in strong turbulence conditions for different values of $\kappa$ and the number of antennas  at the relay $N$.}
	\label{fig:berfixedshad}
\end{minipage}
\end{figure}

Actually,  the  $\cal GK$ fading/shadowing parameters $m$ and $\kappa$ are important and affect the system performance
as shown in Figs.~\ref{fig:berfixed} and \ref{fig:berfixedshad}, respectively. We can see that, heavy shadowing (i.e., small $\kappa$) and/or severe fading (i.e., small $m$) are detrimental for  the system performance.   In Fig.~\ref{fig:berfixedshad}, we fix $\alpha=2.4$, $\beta=2$, $\xi=6.8$, and $r=2$.  Expect for $\kappa=0.6$, we notice that all curves  have the same slopes thereby
inferring that they have the same diversity order. This is due to the fact that the system diversity order is dependent on  ${\rm G}_d=\min\left(N m, \kappa, \frac{\xi^2}{r},\frac{\alpha}{r},\frac{k}{r}\right)$. For the two curves when $\kappa=0.6$, they have the same slope revealing equal diversity order $d=\kappa$. Figs.~\ref{fig:berfixed} and ~\ref{fig:berfixedshad} also show that the asymptotic expansion in  (\ref{eqC2:berfix1high}) agrees very well with the simulation results, hence corroborating its accuracy.

The impact of the  number of relay antennas  $N$ on the system BER is
investigated in Fig.~\ref{fig:berfixedshad} under several shadowing conditions.
As shown in (\ref{div}),  spatial diversity
resulting from employing a higher number of antennas  $N$ at the relay enhances the overall system
performance.
%
\begin{figure}[!t]
	\centering
	\begin{minipage}{.6\textwidth}
		\includegraphics[width=\textwidth]{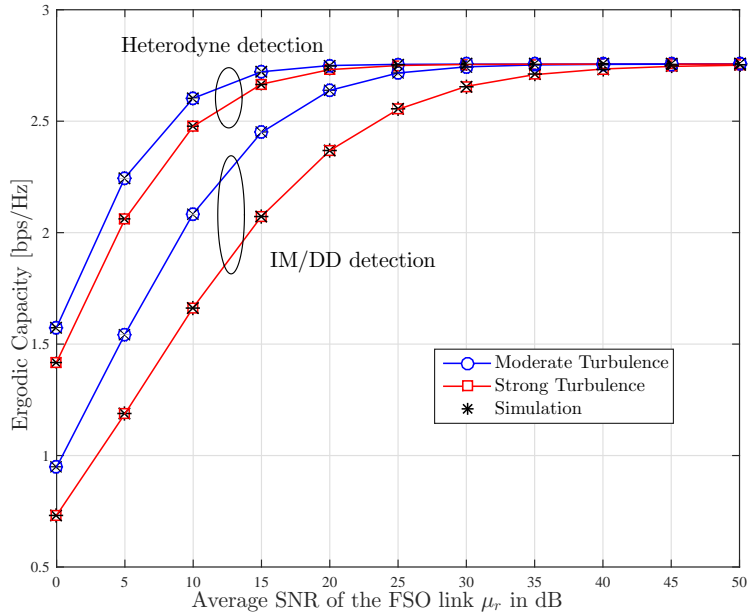}
		\caption{Ergodic capacity of an interference-limited  fixed-gain mixed FSO/RF system in strong and weak turbulence conditions with $N=L=2$, $m=2.5$, and $\kappa=1.09$.}
		\label{fig:capfixed}
	\end{minipage}
\end{figure}

Fig.~\ref{fig:capfixed} shows the impact of the FSO link atmospheric turbulence
conditions on system capacity. We can see that that decreasing
$\alpha$ and $\beta$ (i.e., stronger turbulence conditions) deteriorates the system capacity, notably when IM/DD is employed. It is clear from this figure that   weaker turbulence conditions leads to the situation where the RF link dominates the system performance thereby inhibiting any performance improvement coming from the FSO link.

	
%
	  		  	 	\begin{figure}[t!]
	  	 		\centering
\begin{minipage}{.6\textwidth}
	  	 		\includegraphics[width=\textwidth]{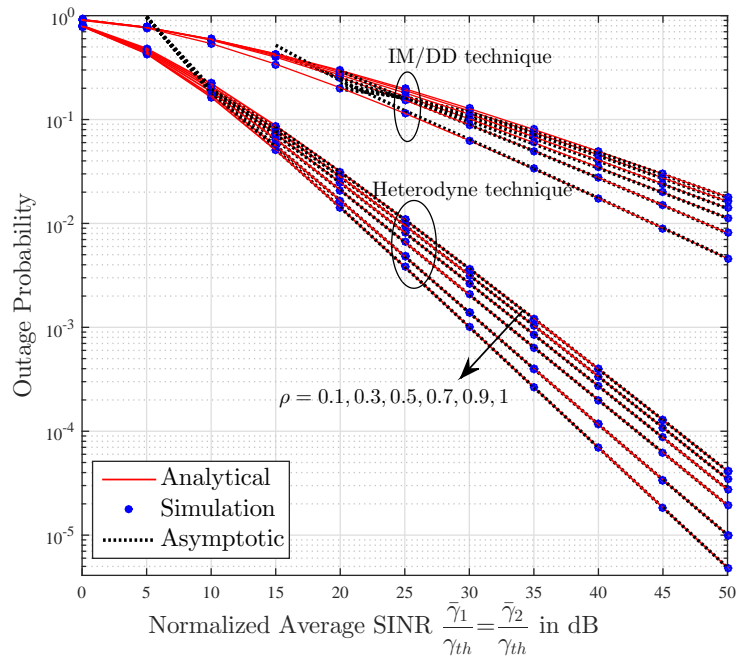}
	  	 		\caption{Outage probability of an interference-limited  CSI-assisted  mixed RF/FSO system for different values of $\rho$ under both detection techniques.}
	  	 		\label{fig:outcsirho}
\end{minipage}
	  	 	\end{figure}
	  	 	
	  	 	Fig. \ref{fig:outcsirho} illustrates the effect of the atmospheric turbulence induced fading severity in terms of the power amount coupled to the LOS component in the FSO link, $\rho$, on the performance of CSI-assisted relay mixed FSO/RF systems. Expectedly, as  $\rho$ increases, the system performance  ameliorates due to the reduction of the atmospheric turbulence over the FSO link. We highlight once again the efficiency of the heterodyne detection against the IM/DD technique.
%
	  	\begin{figure}[t!]
	  		\centering\begin{minipage}{.6\textwidth}
	  		\includegraphics[width=\textwidth]{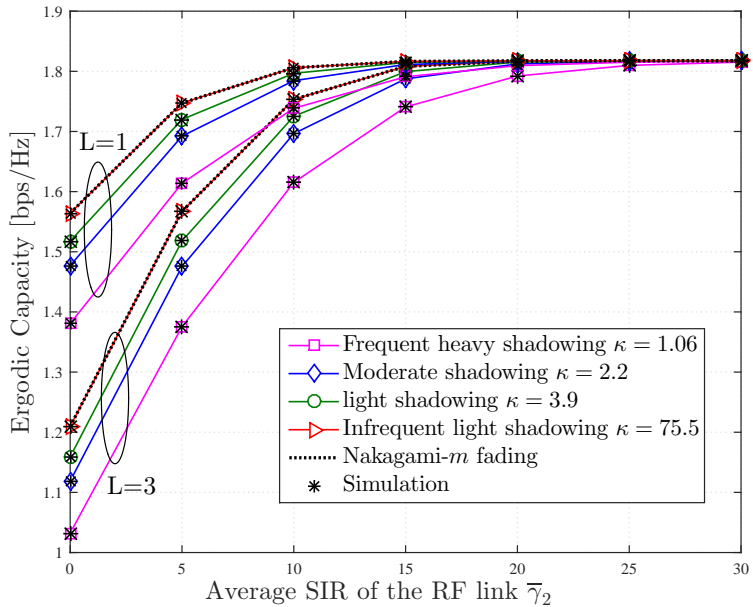}
	  		\caption{Ergodic capacity of an interference-limited  CSI-assisted  mixed FSO/RF relay system in heavy, moderate, and light shadowing for different values of  $L$. }
	  		\label{fig:12}\end{minipage}
	  	\end{figure}   		
	  	
	  	Fig.~\ref{fig:12} investigates the effect of shadowing severity  on the ergodic capacity of mixed FSO/RF  CSI-assisted relaying suffering $\cal GK$ interference. A general observation is that the shadowing degrades the system's overall performance. Furthermore, more interference (i.e., higher $L$) at the RF user results a lower capacity. A similar behavior has been noticed in \cite{imen2}. It may be also useful to mention that the ergodic capacity curves of mixed FSO/RF under infrequent light shadowing  and mixed M\'alaga-$\mathcal{M}$/Nakagami-$m$ systems coincide thereby, unambiguously, corroborating the much wider scope claimed by our novel  analysis framework and the rigor of its mathematical derivations.
	  	
\section{Conclusion}
\label{sec:5}
We have studied  the performance of relay-assisted
mixed FSO/RF  system with RF interference and two different detection techniques.
The {\rm H}-transform theory is involved into a unified performance analysis framework featuring  closed-form expressions for the
outage probability, the BER, and
the channel capacity assuming M\'alaga-$\mathcal{M}$/composite fading/shadowing $\cal GK$
channel models for the FSO/RF links while taking into account pointing errors. The
end-to-end performance of mixed Gamma-Gamma/interference-limited Nakagami-$m$
systems can be obtained as a special case of our results. The latter show that the system diversity order  is related to the the minimum value of the atmospheric turbulence, small-scale fading, shadowing  and pointing error parameters.
\appendices
\renewcommand{\thesection}{\Alph{section}}
\section{CDF of the End-to-End SINR}
\label{app:A}
The CDF of the end-to-end SINR $\gamma$ with fixed-gain relaying scheme can be derived, using \cite[Eq.(8)]{imenefixed} as
\begin{equation}
	\label{eqC2:apA1}
	F_\gamma(x)=\displaystyle \int_{0}^{\infty} F_{\gamma_1}\left( x\left( \frac{{\cal C}}{y}+1\right) \right) f_{\gamma_2}(y)\mathrm{d}y,
\end{equation}		
where  $F_{\gamma_1}$ and $f_{\gamma_2}$ are the FSO link's CDF and the RF link's PDF, respectively. $f_{\gamma_2}$ is derived by differentiation of (\ref{eqC2:5}) over $x$ as\vspace{-0.2cm}
\begin{eqnarray}
	\label{eqC2:PDFRF}
	f_{\gamma_2}(x)&=&\frac{-\kappa m}{\Gamma(Nm)\Gamma(\kappa)\Gamma(Lm_I)\Gamma(\kappa_I)\kappa_I m_I \bar{\gamma}_2}{\rm G}_{4,4}^{3,3}\Biggl[\frac{\kappa mx}{\kappa_Im_I\bar{\gamma}_2}\Bigg\vert  {-1,-\kappa_I,-Lm_I,0 \atop -1,\kappa-1,Nm-1,0}\!\Biggr].
\end{eqnarray}
Substituting (\ref{eq:9}) and (\ref{eqC2:PDFRF}) into (\ref{eqC2:apA1}) while resorting to the integral representation of the Fox-H \cite[Eq.(1.2)]{mathai} and Meijer-G \cite[Eq.(9.301)]{grad} functions yields
\begin{eqnarray}
	\label{eqC2:14}
	F_\gamma(x)&=&\frac{-\xi^2Ar\kappa m }{\Gamma(\alpha)\Gamma(Nm)\Gamma(\kappa)\Gamma(Lm_I)\Gamma(\kappa_I)\kappa_I m_I \bar{\gamma}_2}\sum_{k=1}^{\beta}\frac{b_k}{\Gamma(k)}\frac{1}{4\pi^2i^2}\nonumber \\ && \int_{\mathcal{C}_1}^{}\int_{\mathcal{C}_2}^{}\frac{\Gamma(\xi^2+rs)\Gamma(k+rs)\Gamma(\alpha+rs)}{\Gamma(\xi^2+1+rs)\Gamma(1-rs)}
	 \frac{\Gamma(-rs)\Gamma(-1-t)}{\Gamma(1+t)} \frac{\Gamma(\kappa-1-t)\Gamma(N m-1-t)}{\Gamma(-t)}\nonumber \\ && \Gamma(2+t)\Gamma(1+\kappa_I+t)\Gamma(1+Lm_I+t) \left(\frac{\kappa m}{\kappa_Im_I\bar{\gamma}_2}\right)^{t} \left(\frac{B^r x}{\mu_r}\right)^{-s}\nonumber \\ &&\int_{0}^{\infty}\left( 1+\frac{\cal C}{y}\right) ^{-s}y^{t}\mathrm{d}y\mathrm{d}s\mathrm{d}t,
\end{eqnarray}
where  $i^2=-1$, and $\mathcal{C}_1$ and $\mathcal{C}_2$ denote the $s$ and $t$-planes, respectively. Finally, simplifying $\int_{0}^{\infty}\left( 1+\frac{\cal C}{y}\right) ^{-s}y^{t}\mathrm{d}y$ to $\frac{{\cal C}^{1+t}\Gamma(-1-t)\Gamma(1+t+s)}{\Gamma(s)}$ by means of \cite[Eqs (8.380.3) and (8.384.1)]{grad}  while utilizing the relations $\Gamma(1-rs)=-rs\Gamma(-rs)$, and $s\Gamma(s)=\Gamma(1+s)$ then \cite[Eq.(1.1)]{mittal} yield  (\ref{eqC2:CDFfix}).
\section{Ergodic Capacity under CSI-Assisted Relaying Scheme}
\label{app:B}
From \cite{imen2}, the ergodic capacity can be computed as
\begin{equation}
\label{eq1:7}
C=\frac{1}{2\ln (2)} \displaystyle\int_0^\infty se^{-s}M^{(c)}_{\gamma_1}(s)M^{(c)}_{\gamma_2}(s)ds,
\end{equation}	
where $M^{(c)}_X(s)=\int_{0}^{\infty}e^{-sx}F^{(c)}_X(x) dx$ stands for the complementary MGF (CMGF).
The CMGF of the first hop's SNR $\gamma_1$ under M\'alaga-$\mathcal{M}$ distribution with pointing errors  is given by \cite[Eq.(9)]{wcl}
\begin{equation}
\label{eqC2:apD1}
M^{(c)}_{\gamma_1}(s)=\frac{\xi^2A  r  \mu_r}{\Gamma(\alpha)B^r}\sum_{k=1}^{\beta} \frac{b_k}{\Gamma(k)}{\rm H}_{4,3}^{1,4} \Biggl[\!\frac{\mu_r}{B^r} s\Bigg\vert \  {(\delta_2,\Delta_2)\atop (\lambda_2,\Lambda_2)}\Biggr].
\end{equation}
Moreover, the Laplace transform of the RF link's CCDF yields its CMGF after resorting to \cite[Eq.(7.813.1)]{grad} and \cite[Eq.(1.111)]{mathai} as
\begin{equation}
\label{eqC2:apD2}
M^{(c)}_{\gamma_2}(s)=\frac{s^{-1}}{\Gamma(Nm)\Gamma(\kappa)\Gamma(Lm_I)\Gamma(\kappa_I)}{\rm H}_{3,4}^{3,3}\Biggl[\frac{\kappa_Im_I\bar{\gamma}_2}{\kappa m} s\Bigg\vert\  {(\chi_2,X_2)\atop(\upsilon_2,\Upsilon_2)}\Biggr] .
\end{equation}
Finally, the ergodic capacity expression in (\ref{eq:18}) follows after plugging (\ref{eqC2:apD1}) and (\ref{eqC2:apD2}) into (\ref{eq1:7}) and applying \cite[Eq.(2.2)]{mittal}.

\end{document}